\documentstyle[prl,aps,epsf]{revtex}
\begin{document}
\title{Staggered-spin contribution to nuclear spin-lattice relaxation
in two-leg antiferromagnetic spin-1/2 ladders.}

\author{D. A. Ivanov and Patrick A. Lee}
\address{Department of Physics,
Massachusetts Institute of Technology,
Cambridge, Massachusetts 02139, USA}
\date{July 13, 1998}

\maketitle

\begin{abstract}
We study the nuclear spin-lattice relaxation rate $1/T_1$ in the
two-leg antiferromagnetic spin-1/2 Heisenberg ladder. More
specifically, we consider the contribution to $1/T_1$ from
the processes with momentum transfer $(\pi,\pi)$. In the limit of
weak coupling between the two chains, this
contribution is of activation type with gap $2\Delta$ at low temperatures
($\Delta$ is the spin gap), but crosses over to a slowly-decaying
temperature dependence at the crossover temperature $T\approx\Delta$.
This crossover possibly explains the recent high-temperature
NMR results on ladder-containing cuprates by T.~Imai et al.

\end{abstract}

Recent NMR experiments on Cu$_2$O$_3$ ladders in
A$_{14}$Cu$_{24}$O$_{41}$ compounds 
(A$_{14}$=La$_6$Ca$_8$, Sr$_{14}$, Sr$_{11}$Ca$_3$)
reveal unexpected behavior of the nuclear spin-lattice relaxation rate
$1/T_1$ at temperatures of order of the spin gap \cite{imai}.
While both $^{63}1/T_1$ and $^{17}1/T_1$ exhibit activation-type behavior
at low temperature, a well-pronounced crossover is observed in
$^{63}$Cu nuclear spin relaxation rate. There exist numerous theoretical
studies of $1/T_1$ in antiferromagnetic two-leg ladders
\cite{fukuyama,dalme,sandvik,troyer}, 
but none of them predicts a similar crossover. 
In this paper we restrict  our attention to the undoped spin ladder and
argue that the observed crossover is due to the processes with the
momentum transfer $q=(\pi,\pi)$.

We describe the spin ladder by the Heisenberg Hamiltonian
\begin{equation}
H=\sum_n \left[ J ({\vec S}_n^{(I)} {\vec S}_{n+1}^{(I)}
+ {\vec S}_n^{(II)}{\vec S}_{n+1}^{(II)}) + J_\perp
{\vec S}_n^{(I)} {\vec S}_n^{(II)} \right].
\label{hamiltonian}
\end{equation}
Here $n$ is an integer labeling the rungs of the ladder, the superscripts
refer to the two chains. $J$ and $J_\perp$ are the coupling constants.
From analyzing experimental data on magnetic succeptibility
\cite{johnston} and on $^{17}$O NMR Knight shift \cite{imai}, 
it has been suggested that in Cu$_2$O$_3$ ladders $J_\perp /J \approx 0.5$.

Both the strong-coupling ($J_\perp/J \gg 1$) and the weak-coupling
($J_\perp/J \ll 1$) approaches confirm that the low-lying magnetic
excitations in the ladder are spin-1 magnons with the minimal gap
at $k=(\pi,\pi)$; in the physically
relevant range $J_\perp /J \approx 0.5$, the value of the
gap is known to be $\Delta \approx 0.5 J_\perp$ \cite{barnes,troyer,shelton}.
We believe that the weak-coupling limit captures the characteristic 
features of the nuclear spin  relaxation. For the sake of simplicity
we assume that $T,J_\perp \ll J$ (with no assumptions on the relative
magnitudes of $J_\perp$ and the temperature $T$). Then all excited magnons
have momenta close to the wavevector $k=(\pi,\pi)$. The relaxation
rate $1/T_1$ is given by
\begin{equation}
{1\over T_1}= 2 \sum_q F(q) S(q,\omega_0)
=\sum_q F(q) {2 T\over\omega_0} {\rm Im} \chi(q,\omega_0),
\label{rrate}
\end{equation}
where $F(q)$ are the appropriate coupling constants, $\omega_0$
is the nuclear resonance frequency ($\omega_0\ll\Delta,T$), 
$S(q,\omega)$ is the dynamical
structure factor, and $\chi(q,\omega)$ is the dynamical magnetic 
succeptibility. We employ the system of units with $\hbar=k_B=\mu_B=1$.
Since the excited magnons have momenta close to $(\pi,\pi)$, there are
two major contributions to $1/T_1$: that with $q\approx (0,0)$ (via
even-magnon-number processes) and that with $q\approx (\pi,\pi)$
(via odd-magnon-number processes). Accordingly, define
\begin{equation}
\left( {1\over T_1} \right)_{q=0}
=2 \int_{q\approx 0} {dq\over 2\pi} S((q,0), \omega_0),
\qquad
\left( {1\over T_1} \right)_{q=\pi}
=2 \int_{q\approx \pi} {dq\over 2\pi} S((q,\pi), \omega_0)
\label{definition}
\end{equation}
(defined this way, the quantities $(1/T_1)_{q=0}$ and $(1/T_1)_{q=\pi}$
do not have the dimension of inverse time).
Then the total relaxation rate $1/T_1$ is a linear combination of
$(1/T_1)_{q=0}$ and $(1/T_1)_{q=\pi}$.

The coupling constants $F(q)$ for the oxygen sites vanish at 
$q=(\pi,\pi)$ \cite{imai}, so
\begin{equation}
^{17}{1\over T_1} \propto \left( {1\over T_1} \right)_{q=0}.
\end{equation}
The experiments on SrCu$_2$O$_3$ suggest that in Cu$_2$O$_3$ ladders
the coupling constant $F(q)$ for the copper site is dominated by the
on-site hyperfine interaction and therefore only weakly depends
on $q$ \cite{ishida}. Thus we expect that $^{63}1/T_1$ is a 
linear combination of $(1/T_1)_{q=0}$ and $(1/T_1)_{q=\pi}$ with
the coefficients of the same order of magnitude. Next, the available numerical
studies\cite{sandvik} indicate that at temperatures of order
of the spin gap the contributions
$(1/T_1)_{q=0}$ and $(1/T_1)_{q=\pi}$ have comparable
magnitudes. Therefore, we suggest that $^{63}1/T_1$ in the
experimentally relevant temperature range is not dominated
by the $q=0$ contribution, in contrast with other studies
\cite{fukuyama,dalme}.

A simple argument shows that the contribution $(1/T_1)_{q=\pi}$
has gap $2\Delta$ as opposed to gap $\Delta$ for $(1/T_1)_{q=0}$
(this argument was originally proposed for spin-1 chains \cite{sagi}, but
it also remains valid for spin-1/2 ladders).
In  $(1/T_1)_{q=0}$, the nuclear spin is relaxed by the quasi-elastic
scattering of a thermally excited magnon, which requires a gap of $\Delta$.
On the other hand, $(1/T_1)_{q=\pi}$ assumes an elastic scattering
with momentum transfer near $(\pi,\pi)$, which means that the
total number of incoming and outcoming magnons must be odd (each magnon
carries momentum close to $(\pi,\pi)$). Processes of creating and
annihilating a single magnon cannot occur at zero energy because
of the gap. Therefore, 
the leading contribution comes from three-magnon processes:
two thermally excited magnons are converted into a single magnon
carrying the total energy of the incoming magnons (up to the NMR
frequency $\omega_0$), or vice versa
(see Fig.~1 below). Such processes require energy at least $2\Delta$.

On the basis of the above argument, the contribution 
$(1/T_1)_{q=\pi}$ was usually neglected. However,
several recent works conclude that the effective gap in $(1/T_1)_{q=0}$
is somewhat larger than $\Delta$ (determined from succeptibility),
either because of magnon interaction \cite{dalme} or due to the singlet
excitation mode \cite{fukuyama}. A larger gap in $1/T_1$ than in 
the succeptibility is indeed reported in Cu$_2$O$_3$ ladders
\cite{azuma,ishida}. In view of these results, our suggestion of importance
of $q=\pi$ contribution to $^{63}$Cu nuclear spin relaxation rate
appears more plausible.

For the rest of the paper we focus on computing $(1/T_1)_{q=\pi}$
in the weak-coupling limit $J_\perp \ll J$.
Although we are unable to find a closed analytic form for $(1/T_1)_{q=\pi}$
in the whole range of temperatures, we 
find the high- and low-temperature asymptotics and estimate 
the crossover temperature.

According to the results of \cite{shelton}, in the weak-coupling limit,
the spin-1/2 two-leg ladder is equivalent to four massive Majorana fermions,
combined into a triplet of mass $\Delta$ and a singlet of mass $3\Delta$:
\begin{equation}
H_f=H_\Delta[\xi_1] + H_\Delta[\xi_2] + H_\Delta[\xi_3] + H_{3\Delta}[\rho]
+ H_{int},
\end{equation}
where each $H_m[\xi]$ is a free massive Hamiltonian of a Majorana fermion
$\xi$;  $H_{int}$ is the four-fermion interaction arising from the marginal
term in the interchain coupling \cite{shelton}. It has been
argued in \cite{dalme} that the interaction would lead to nonperturbative
effects in $(1/T_1)_{q=0}$. However, the staggered magnetization is 
nonlocal in terms of fermions, and we expect that the interaction will
 play a smaller role in $(1/T_1)_{q=\pi}$. Following \cite{shelton},
we neglect $H_{int}$. 

In the continuum limit, the local magnetization 
$\vec S^{(\alpha)}$ may be split into the
uniform and staggered components:
\begin{equation}
\vec S^{(\alpha)}_n = \vec J^{(\alpha)} + (-1)^n \vec n^{(\alpha)}.
\end{equation}
While $\vec J$ has a quadratic expression in terms of the Majorana
fermions, $\vec n$ is non-local in the fermion operators. If the Majorana
fermions are mapped onto non-critical 1+1-dimensional 
Ising models (we need four Ising
models --- one for each Majorana fermion), the operator $\vec n$ may
be expressed in terms of order ($\sigma$) and disorder ($\mu$) parameters
of the Ising models \cite{shelton}. In particular,
\begin{equation}
n^-_z = n^{(I)}_z - n^{(II)}_z \propto \sigma_1 \sigma_2 \mu_3 \sigma^*,
\label{staggered}
\end{equation}
where $\sigma^*$ is the order parameter for the Ising model with gap 
$\Delta^*=3\Delta$;
$\sigma_1$, $\sigma_2$, and $\mu_3$ --- order and disorder parameters of the
three identical Ising models with gap $\Delta$ \cite{shelton}.
Conventionally, we have chosen the Ising models 
to be in ordered state, so that
$\langle \sigma \rangle \ne 0 $ at zero temperature.

The expression (\ref{staggered}) would enable us to compute $(1/T_1)_{q=\pi}$
using (\ref{definition}),
should we know the non-critical Ising correlation functions at finite 
temperature. In a recent work \cite{leclair}, the latter has been 
expressed as a series in number of fermionic excitations, and we shall use
their result to match the high- and low-temperature asymptotics of the
Ising correlation functions.

The prportionality coefficient in (\ref{staggered}) is non-universal.
To fix the relative normalization of the high- and low-temperature
asymptotics, define
\begin{equation}
\tilde n^-_z = \sigma_1 \sigma_2 \mu_3 \sigma^*
\label{norm1}
\end{equation}
with the order and disorder operators normalized by their short-distance
asymptotics:
\begin{equation}
\langle \sigma(x) \sigma(0) \rangle \sim
\langle \mu(x) \mu(0) \rangle \sim
{1\over |x|^{1/4}} , \qquad x\to 0.
\label{norm2}
\end{equation}
In what follows we normalize $(1/T_1)_{q=\pi}$ accordingly, in other
words, in (\ref{definition}) we set $S(x,t)=\langle 
\tilde n^-_z (x,t) \tilde n^-_z (0,0) \rangle$.

In a natural way, the following three limits may be distinguished:

(i) $T\ll\Delta$. In this low-temperature limit, $(1/T_1)_{q=\pi}$ has  
activation-type behavior with gap $2\Delta$. The prefactor may be computed
using the results of \cite{leclair}.

(ii)  $T\gg\Delta^*$. In this limit the gaps presumably play no role,
and the system is equivalent to two uncoupled chains (``quantum critical''
phase). The result for $(1/T_1)_{q=\pi}$ in this limit may be borrowed
from \cite{sachdev}. 
Although this high temperatures are beyond the experimental range,
considering this limit is useful for determining the degree of applicability
of the approximations being made.

(iii) $\Delta \ll T \ll \Delta^*$. In the ladder system this intermediate
limit is never attained, since $\Delta^*=3\Delta$. However we may formally
take this limit assuming $\Delta^* \gg \Delta$. The validity of this
approximation will be discussed further.

Below we describe in more detail the computations and the results in these
three limits.

(i) $T\ll\Delta$. According to \cite{leclair}, the (non-ordered) Ising
correlation functions may be expressed as the series in the number of fermionic 
excitations:
\begin{equation}
\langle \sigma(t,0) \sigma(0,0) \rangle = \sigma_0^2
\sum_{n {\rm ~even}} {1\over n!} \sum_{\varepsilon_i}
\int_{-\infty}^{+\infty} \prod_{i=1}^n {d\beta_i \over 2\pi}
f_{\varepsilon_i}(\beta_i) e^{-i \varepsilon_i E(\beta_i) t}
| F(\beta_1,\dots,\beta_n)_{\varepsilon_1,\dots,\varepsilon_n}|^2,
\label{series}
\end{equation}
where $\varepsilon_i=\pm 1$ ($i=1,\dots,n$) are the particle-hole indices,
\begin{equation}
f_{\varepsilon}(\beta)=\left[ 1+\exp(-\varepsilon E(\beta)/T) \right]^{-1}
\end{equation}
is the Fermi distribution function, $\beta_i$ are the rapidities of the
excitations with energies $E(\beta)=\Delta \cosh \beta$.
\begin{equation}
F(\beta_1,\dots,\beta_n)_{\varepsilon_1,\dots,\varepsilon_n}
=i^{n/2} \prod_{i<j} \left( \tanh {\beta_i-\beta_j \over 2} 
\right)^{\varepsilon_i \varepsilon_j}
\label{formfactors}
\end{equation}
are the corresponding formfactors \cite{leclair,yurov}.

The same expression gives $\langle \mu(t,0) \mu(0,0) \rangle$,
with the only difference that the sum is taken over odd numbers of
excitations $n$.

The zero-temperature magnetization $\sigma_0$ may be taken from the
exact result on the Ising model \cite{wu,typo}:
\begin{equation}
\sigma_0^2= \Delta^{1/4} 2^{1/6} e^{-1/4} A^3\approx 1.8437\Delta^{1/4},
\label{sigma0}
\end{equation}
where $A=\exp[1/12 - \zeta^\prime(-1)]=1.282427\dots$ is the Glaisher 
constant. In the Ising model with gap $3\Delta$, 
${\sigma^*_0}^2=3^{1/4} \sigma_0^2$.

Using (\ref{series}), we express
\begin{equation}
\left({1\over T_1}\right)_{q=\pi} = 2\int_{-\infty}^{+\infty}
dt \langle \tilde n^-_z (x,t) \tilde n^-_z (0,0) \rangle
\label{def2}
\end{equation}
as a sum of the odd-magnon-number processes with zero energy
transfer. The one-magnon process has a gap and does not contribute
to (\ref{def2}). In Fig.~1 we show the three types of three-magnon
processes contributing to $(1/T_1)_{q=\pi}$.


\begin{figure}
\centerline{\epsffile{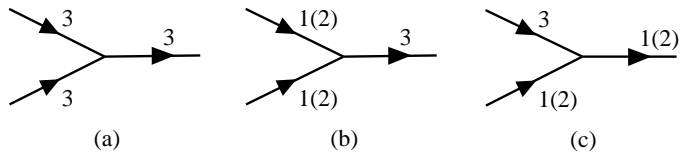}}
\caption{Three-magnon processes contributing to $(1/T_1)_{q=\pi}$.}
\label{fig1}
\end{figure}

In the diagrams, the arrows correspond to the particle-hole
index $\varepsilon_i$ in (\ref{series}), the labels 1,2,3 --- to the three
different Majorana fermions (equivalently, to the spin of magnons). The
contribution from the diagrams shown in Fig.~1 must be multiplied by two
to account for the diagrams with arrows reversed. We neglect processes
including the singlet channel (with gap $3\Delta$), since they have a larger
gap $4\Delta$. The processes (a) and (b) in Fig.~1 can be shown to have
temperature dependence $T^2 \exp (-2\Delta/T)$, while the process (c) 
dominates at low temperatures with the temperature dependence 
$T \exp (-2\Delta/T)$. Including only the latter contribution, we find:
\begin{equation}
\left({1\over T_1}\right)_{q=\pi} = 4 \sigma_0^6 {\sigma_0^*}^2 
\int\!\!\!\!\!\int\limits_{-\infty}^{+\infty}\!\!\!\!\!\int
{d\beta_1\over 2\pi}
{d\beta_2\over 2\pi}
{d\beta_3\over 2\pi}
{2\pi\delta\left(E(\beta_1)+E(\beta_2)-E(\beta_3)\right)  \over
8\cosh{E(\beta_1)\over 2T}
\cosh{E(\beta_2)\over 2T}
\cosh{E(\beta_3)\over 2T}}
\coth^2\left({\beta_1-\beta_3\over 2}\right).
\label{lowT1}
\end{equation}
The low-T asymptotics of this expression is 
\begin{equation}
\left({1\over T_1}\right)_{q=\pi} \sim {4\sqrt{3}\over \pi}
\left({\sigma_0^6 {\sigma_0^*}^2 \over \Delta}\right)
\left({T\over\Delta}\right) e^{-{2\Delta\over T}}.
\label{lowT2}
\end{equation}
Putting in the numbers,
\begin{equation}
\left({1\over T_1}\right)_{q=\pi} \sim 33.54
\left({T\over\Delta}\right) e^{-{2\Delta\over T}}
\label{lowT3}
\end{equation}
(with the normalization (\ref{norm2}), $(1/T_1)_{q=\pi}$
is dimensionless).

(ii) In the high-temperature limit $T\gg\Delta^*$, the quantity
$(1/T_1)_{q=\pi}$ is determined by the short-distance asymptotics of
the correlation function
\begin{equation}
\langle \tilde n^-_z (x) \tilde n^-_z (0) \rangle \sim {1\over x}.
\end{equation}
In this case Sachdev's result \cite{sachdev} gives
\begin{equation}
(1/T_1)_{q=\pi} = \pi \approx 3.1416.
\label{highT3}
\end{equation}

(iii) In this limit, we replace the operator $\sigma^*$ by its
zero-temperature expectation value, while pretending that the three remaining
operators in (\ref{norm1}) are massless. Then
\begin{equation}
\langle \tilde n^-_z (x) \tilde n^-_z (0) \rangle \sim 
{{\sigma_0^*}^2\over x^{3/4}}.
\label{power34}
\end{equation}

To handle this case, we redo the calculation of \cite{sachdev} for
arbitrary exponent $\eta$ in
\begin{equation}
\langle \tilde n^-_z (x) \tilde n^-_z (0) \rangle \sim 
{D\over x^\eta}.
\end{equation}
The magnetic succeptibility is then given by \cite{shankar}
\begin{equation}
\chi(\omega,k)= {\pi D \over (2\pi T)^{2-\eta}}
{\Gamma\left(1-{\eta\over2}\right)\over
\Gamma\left({\eta\over2}\right)}
{\Gamma\left({\eta\over4}-i{\omega+k\over4\pi T}\right)
\Gamma\left({\eta\over4}-i{\omega-k\over4\pi T} \right)\over
\Gamma\left(1-{\eta\over4}-i{\omega+k\over4\pi T}\right)
\Gamma\left(1-{\eta\over4}-i{\omega-k\over4\pi T} \right)},
\end{equation}
which leads to
\begin{equation}
\left({1\over T_1}\right)_{q=\pi}=\lim_{\omega\to 0}
\int_{-\infty}^{+\infty}
{dk\over 2\pi}
{2T\over\omega}
{\rm Im}\chi(\omega,k) =
D{\Gamma^2\left({\eta\over2}\right)\over
\Gamma(\eta)}
(2\pi T)^{\eta-1}.
\end{equation}
In our case ($\eta=3/4$, $D={\sigma_0^*}^2$),
\begin{equation}
\left({1\over T_1}\right)_{q=\pi}=
{\sigma_0^*}^2{\Gamma^2\left({3\over8}\right)\over
\Gamma\left({3\over4}\right)}
(2\pi T)^{-1/4}.
\label{medT2}
\end{equation}
Numerically,
\begin{equation}
\left({1\over T_1}\right)_{q=\pi}=
7.028 \left({T\over\Delta}\right)^{-1/4}.
\label{medT3}
\end{equation}

The three temperature dependences (\ref{lowT3}), (\ref{highT3}),
and (\ref{medT3}) are plotted in Fig.~2.

The mismatch of the asymptotics (ii) and (iii) at $T\approx 3\Delta$
is due to the roughness of the approximation 
$\langle \sigma^*(t) \sigma^*(0) \rangle = {\sigma^*_0}^2 $ made
in  (\ref{power34}). At $T\approx\Delta^*$, this correlation function
decays at time scale of order $T^{-1}$ (the same as the triplet-channel
correlation functions), which substantially decreases $(1/T_1)_{q=\pi}$.
Although for $\Delta^*=3\Delta$ the limit (iii) is never realized,
it clearly indicates that at high temperature $(1/T_1)_{q=\pi}$
slowly decreases with temperature, approaching the constant value of the 
asymptotics (ii).


\begin{figure}
\epsfxsize=3in
\centerline{\epsffile{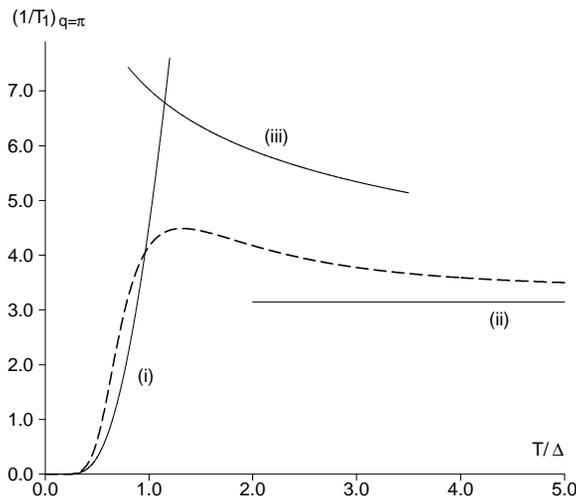}}
\caption{The three limiting asymptotics of the temperature
dependence of $(1/T_1)_{q=\pi}$. Dashed line is the qualitative
interpolation of the actual temerature dependence.}
\label{fig2}
\end{figure}

Another qualitative consequence of our discussion is the crossover from
increasing activation behavior of asymptotics (i) to slowly decreasing
high-temperature asymptotics (iii)--(ii). The crossover occurs at
$T\approx\Delta$, in spite of the gap $2\Delta$ of the low-temperature
asymptotics. This agrees with the experimental results
of T.~Imai et~al., who observed a sharp corssover in $^{63}(1/T_1)$ at
$T\approx 425$K in the undoped compound La$_6$Ca$_8$Cu$_{24}$O$_{41}$
\cite{imai}. Moreover, the form of the temperature dependence of 
$^{63}(1/T_1)$ in the hole-doped compounds Sr$_{14}$Cu$_{24}$O$_{41}$ and 
Sr$_{11}$Ca$_3$Cu$_{24}$O$_{41}$ appear qualitatively very close to
the proposed form of $(1/T_1)_{q=\pi}$ (including a slowly-decreasing high-$T$
behavior). We may speculate that this can possibly be explained by 
a different hyperfine coupling of the $^{63}$Cu spin in these compounds
making $(1/T_1)_{q=\pi}$ contribution dominate in $^{63}(1/T_1)$. On the
other hand, hole doping may modify the temperature dependence
of $(1/T_1)_{q=\pi}$, which deserves further theoretical study.

It is worth mentioning that our interpolation of $(1/T_1)_{q=\pi}$
(dashed line in Fig.~2) is higher than the asympthotics (i) at low
temperature. This comes from the fact that all terms in the
low-temperature expansion (\ref{series}) are positive, and including
only the leading term underestimates $(1/T_1)_{q=\pi}$ at low
temperatures. This observation ensures that the crossover from (i)
to (iii)-(ii) is sufficiently sharp.

To experimentally separate $(1/T_1)_{q=\pi}$ and $(1/T_1)_{q=0}$
contributions in the total spin-lattice relaxation rate
$^{63}(1/T_1)$, one can use the fact that
$q=\pi$ contribution does not have a singular dependence on the
external field. While
$(1/T_1)_{q=0}$ singularly diverges at low magnetic fields (as
$\log H$ in the free-magnon model \cite{troyer} or as $H^{-1/2}$
in the spin-diffusion model \cite{dalme,takigawa}), $(1/T_1)_{q=\pi}$
only weakly depends on the magnitude of the magnetic field.

We hope that further experimental and numerical works will provide a better
understanding of the crossover in $(1/T_1)_{q=\pi}$ discussed in this
paper.

The authors would like to thank T.~Imai for sharing the experimental data
prior to publication, A.~M.~Tsvelik for drawing our attention to
ref.~\cite{leclair}, and D.~H.~Kim for many helpful discussions.

This research was supported by the MRSEC program of NSF under
award number DMR 94-00334.


\begin{references}
\bibitem{typo}
In Ref. \cite{yurov}
the citation of the result (\ref{sigma0}) has a typo.
\end{references}
\end{document}